\documentclass[prl,twocolumn,superscriptaddress,showpacs,floatfix]{revtex4}
\usepackage{graphicx}
\begin{document}
\newcommand{\RR}{\mathrm{\mathbf{R}}}
\newcommand{\rr}{\mathrm{\mathbf{r}}}
\newcommand{\defin}{\stackrel{def}{=}}

\title{Single electron spin and its coherence in Si quantum computer architecture}
\author{M.J. Calder\'on}

\affiliation{Condensed Matter Theory Center, Department of Physics,
University of Maryland, College Park, MD 20742-4111}

\author{Belita Koiller}


\affiliation{Instituto de F\'{\i}sica, Universidade Federal do Rio de
Janeiro, Cx. Postal 68528, 21941-972 Rio de Janeiro, Brazil}

\author{S. {Das Sarma}}

\affiliation{Condensed Matter Theory Center, Department of Physics,
University of Maryland, College Park, MD 20742-4111}

\date{\today}



\begin{abstract}
The possibility of performing single spin measurements in Si-based quantum computers through electric field control of electrons bound to double donors near a barrier interface is assessed. We find that both the required electric fields and the tunneling times involved are probably too large for practical implementations. On the other hand, operations with double donors in their first excited state require smaller fields and faster tunneling times, and are therefore suitable for spin-to-charge conversion measurements. We also propose a measurement scheme that would render statistical (ensemble) estimates of the spin coherence at the Si/SiO$_2$ interface.  
\end{abstract}

\maketitle

Among the several operations required for a spin-based quantum computer (QC), single spin rotations and measurement are probably the 
hardest ones to  achieve within the  currently available technology. 
Not only is the electron spin extremely weak to be detected~\cite{Rugar04}, operation and measurement times must be fast enough compared to electronic spin dephasing times.
Single electron spin control has very recently been demonstrated in a double quantum dot (QD) configuration in a GaAs heterostructure by Koppens {\it et al.}~\cite{naturebyKoppens}.
In the occupation number representation for the individual dots, the involved states are (1,1), one electron in each dot, and (0,2), 
one of the dots empty and the other doubly-occupied. Due to Pauli principle, state (0,2) is only accessible if the electrons form a spin singlet, which allows inferring spin states from charge transport measurements~\cite{naturebyKoppens}.
Although a similar experiment has been suggested using double donors
in Si~\cite{Kane00PRB} and other approaches have been
attempted~\cite{kenton06} for single
donor-based QC in Si~\cite{Kane}, progress in Si systems has been much
slower than in the corresponding GaAs quantum dot systems, primarily
because precise quantitative practical schemes for spin-to-charge conversion have
not been theoretically proposed in Si QC architectures.  The proposal by  Kane {\it et al.}~\cite{Kane00PRB} has many of the basic ingredients that led to the recent successful results in GaAs QDs: The double dot potential would correspond to a double well structure formed by a double donor in Si near the interface with a barrier (e.g. Si/SiO$_2$) under a uniform electric field applied perpendicular to the interface.
The (1,1) state would correspond to one electron at the characteristic triangular shaped interface-plus-electric-field  well  and one at the double donor Coulomb well, while the (0,2) state would correspond to both electrons bound to the (neutral) double donor.

We explore here the use of double donors in Si (solid-state analogues of the He atom) to perform one-qubit operations similar  to those reported in Koppens's experiments. For the measurement of a single spin state, we quantitatively estimate the required electric fields for the first ionization of Te in Si, assessing the practical implementation of the scheme  in Ref.~\onlinecite{Kane00PRB}. We also present a related proposal for the measurement of spin coherence times in Si near a SiO$_2$ interface. This is a key parameter if qubit measurement in Si-based QC is to be performed at the interface~\cite{Kane}.
The idea proposed here is similar to the one that has allowed the measurement of spin dephasing times in double quantum dots with two electrons in GaAs~\cite{petta05}. In this case we show that S double donors would be the most appropriate due to the negligible spin-orbit coupling.

\begin{figure}
\begin{center}
\resizebox{58mm}{!}{\includegraphics{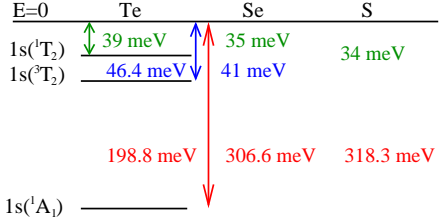}}
\caption{(Color online) First ionization energies of the 1s levels of double donors (Te, Se, S) in Si. The ground state is the singlet 1s($^1$A$_1$)$=$ [1s(A$_1$), 1s(A$_1$)] and the first excited states are 1s($^i$T$_2$)$=$ [1s(A$_1$), 1s(T$_2$)]. ($i=1$) corresponds to a singlet while ($i=3$) to a triplet. Data taken from Refs.~\onlinecite{grimmeis92,bergman88}.\\
}
\label{fig:1s-levels-dd}
\end{center}
\end{figure}

Double donors (S, Se, T) in Si are substitutional deep centers whose electrons' binding energies (summarized in Table~\ref{table:data}) are typically one order of magnitude larger than for single donors (P, As, Sb).
The substitutional double donor ground state is a spin singlet analogous to the He atom ground state.  For comparison, we recall that the two-electrons binding energy in neutral He is $5.7$ Ry, while the one-electron binding energy in He$^+$ is $4$ Ry, thus the first ionization energy in He is $1.7$ Ry~\cite{slater}. 
The particular band structure of Si and the local tetrahedral  symmetry of the potential leads to an orbital ground state for substitutional donors of A$_1$ symmetry. The first excited state for double donors is such that one of the electrons is in the ground state and the other on the next-in-energy 1s(T$_2$) orbital state~\cite{ning71-II}. This outer electron experiences a closely hydrogenic effective potential because the double donor core charge is virtually screened by the inner 1s(A$_1$) electron~\cite{ning71-II}, which has a very small effective Bohr radius. 
The corresponding spin states are always singlet for the ground state, while the first excited configuration can be in a triplet or a singlet, with the triplet lower in energy due to Hund's rule. The binding energies of the ground and the first excited states of S, Se, and Te are illustrated in Fig.~\ref{fig:1s-levels-dd}.

We consider a double donor (S, Se or Te) on its ground state in Si, a distance $d$ away from an interface with SiO$_2$.  When a characteristic electric field $F_{\rm c}$ perpendicular to the interface is applied, one of the electrons may tunnel towards the interface well (where it still feels the Coulomb attraction of the donor which keeps it from spreading in a 2-dimensional electron gas~\cite{QC-PRL06}). Small changes of the electric field around $F_{\rm c}$ cause the electron to move between the donor and the interface. This `shuttling' is only allowed as long as the two electrons are in a spin-singlet state while if they are in a spin-triplet state the electron cannot tunnel back to the donor due to Pauli principle. The electron motion can be detected by single-electron transistors (SETs) located on the device surface, above the oxide layer~\cite{Kane00PRB}. As a result, the measurement of charge directly leads to information about the spin state of the electron (spin-to-charge conversion). 

As shown in Table~\ref{table:data}, Te is the shallowest of the double donors in Si, with a total (two electrons) binding energy $E_{\rm Te}=609.6$ meV. For this reason it is the most apropriate for electric field driven ionization experiments in Si, as it requires the smallest electric field $F^{\rm Te}_{\rm c}$ to ionize. This field has been estimated, within a single electron approach, in Ref.~\onlinecite{Kane00PRB}. Here we consider  the full two-particle Hamiltonian, which includes the electron-electron interaction and the image charges, to calculate this ionization field as a function of $d$.

\begin{table}[t]
\begin{tabular}{|c|c|c|c|}
\hline
Donor  & E$_{\rm D}$ (meV) & E$_{\rm D^+}$ (meV) & 1$^{\rm st}$ ionization (meV) \\
\hline   
Te &609.6  &410.8  & 198.8 \\
\hline
Se &899.9  &593.3  &306.6  \\
\hline
S  & 931.8 & 613.5 & 318.3 \\
\hline  
\end{tabular}
\caption{Energy of neutral (E$_{\rm D}$) and singly ionized (E$_{\rm D^+}$) double donors. The first ionization energy is E$_{\rm D^+}$-E$_{\rm D}$. Data taken from Ref.~\onlinecite{grimmeis92}.}
\label{table:data}
\end{table}

We solve the 2-electron double donor problem within several simplifying approximations, namely single-valley effective mass approach, assuming isotropic effective mass $m^*$.
The validity of these approximations for obtaining ionization fields is discussed below. 
The full two particle Hamiltonian in atomic units ($a^*={{\hbar^2\epsilon_{Si}}/{m^* e^2}}$ and $Ry^*={{m^* e^4}/{2\hbar^2\epsilon_{Si}^2}}$) for a double donor in Si close to a Si/SiO$_2$ (0,0,1) interface in the presence of an electric field is 
\begin{eqnarray}
H&=&T_1+T_2-\frac{4}{r_1}-\frac{4}{r_2}+\kappa eF(z_1+z_2)+\frac{2}{r_{12}}\nonumber \\
&+& V^{\rm D}_{\rm image}+V^{e}_{\rm
    image} ,
\label{eq:hamiltonian}
\end{eqnarray}
with kinetic energy $T_i=- {{\partial^2}/{\partial x_i^2}} - {{\partial^2}/{\partial
 y_i^2}}- {{\partial^2}/{\partial
 z_i^2}} $
where $i=1,2$, $\kappa=3.89 \times 10^{-7} \epsilon_{Si}^3
 \left({{m}/{m^*}}\right)^2$ cm/kV, and the electric field $F$ is
 given in kV/cm. We estimate the isotropic radius $a_{\rm Te}$ and mass $m^*$ from the experimental value of the binding energy of Te in the ground state, $E_{Te}= -609.6 $ meV, using the calculated expression for He, $E_{Te}=2/a_{\rm Te}^2-8/a_{\rm Te}+5/4a_{\rm Te}=-5.6953 Ry^*$ which is minimized by $a_{\rm Te}=0.59a^*$~\cite{slater}.  This gives $Ry^*=107$ meV, $m^* \sim m$, and $a^*=0.6$ nm. The image terms are

\begin{equation}
V^{\rm D}_{\rm image}=\frac{4
  Q}{\sqrt{x_1^2+y_1^2+(z_1+2d)^2}}+\frac{4
  Q}{\sqrt{x_2^2+y_2^2+(z_2+2d)^2}}\,\, ,
\label{eq:imageD}  
\end{equation}  
the interaction of the electrons with the double donor image, and 
\begin{eqnarray}
\label{eq:imagee} 
&V^{e}_{\rm
    image}&=
-\frac{Q}{2(z_1+d)}-\frac{Q}{2(z_2+d)}\\
&-&\frac{4 Q}{\sqrt{(x_1-x_2)^2+(y_1-y_2)^2+(z_1+z_2+2d)^2}} \nonumber\,\, ,
\end{eqnarray}
the interaction of the electrons with their own images and the other electron's image. $Q={{(\epsilon_{\rm SiO_{2}}-\epsilon_{\rm Si})}/{(\epsilon_{\rm
 SiO_{2}}+\epsilon_{\rm Si})}}$,
 where $\epsilon_{\rm Si}=11.4$ and  $\epsilon_{\rm SiO_{2}}=3.8$.

We solve the Hamiltonian in the non-orthogonal basis defined by the initial state  $\Psi_A\equiv(0,2)$ with the two electrons on the donor ground state [1s(A$_1$),1s(A$_1$)]:
\begin{equation}
\Psi_A = \psi_D (1) \psi_D (2)\,\,,
\end{equation}
and the final singly ionized donor state  and one electron at the interface $\Psi_B\equiv(1,1)$
\begin{equation}
\Psi_B =\frac{1}{\sqrt{2 (1+S_{ID}^2)}} \left[\psi_D (1) \psi_I
  (2)+\psi_D (2) \psi_I (1)\right] \,\,,
\end{equation}
where $i=1,~2$, $\psi_D (i) \propto e^{{r_i}/{a_{\rm Te}}}$, $\psi_I(i) \propto 
 (z_i+d)^{2} e^{-\alpha (z_i+d)/2} \times e^{-\beta^2
  \rho_i^2/2}$,
and $S_{ID}=\langle \psi_D |\psi_I \rangle$ is the overlap between the single electron states. The variational parameters $a_{\rm Te}$, $\alpha$ and $\beta$ minimize the expectation value of the energy and lead to approximate expressions for the one-electron ground state $\psi_D$ and $\psi_I$ at the donor well and at the interface well respectively.

Note that the Bohr radius of the double donor ground state ($a_{\rm Te}=0.354$ nm) is one order of magnitude smaller than the one for the single donor ($a_{\rm P}= 2.365$ nm) due to the larger binding energy. This fact allows two useful simplifications to be made. First, the overlap 
$S_{ID}$ is very small, therefore the exchange part of the electron-electron interaction can be neglected. 
Second, in the limit in which $d >> a_{\rm Te}$ the electron at the interface sees a nuclear image with charge $+1$ (H-like atom) rather than a $+2$ donor plus an electron (first ionized He-like atom). Therefore, all the image terms (Eqs.~\ref{eq:imageD} and \ref{eq:imagee}) in the Hamiltonian are reduced to
\begin{eqnarray}
& &V^e_{\rm image}+V^D_{\rm image}\approx -\frac{Q}{2(z_1+d)}-\frac{Q}{2(z_2+d)} \nonumber \\
&+&\frac{2
  Q}{\sqrt{x_1^2+y_1^2+(z_1+2d)^2}} +\frac{2
  Q}{\sqrt{x_2^2+y_2^2+(z_2+2d)^2}}\,\,,
\end{eqnarray} 
and the variational parameters for $\psi_I(i)$ are the same as for the single donor problem~\cite{QC-PRL06}. 

\begin{figure}
\begin{center}
\resizebox{60mm}{!}{\includegraphics{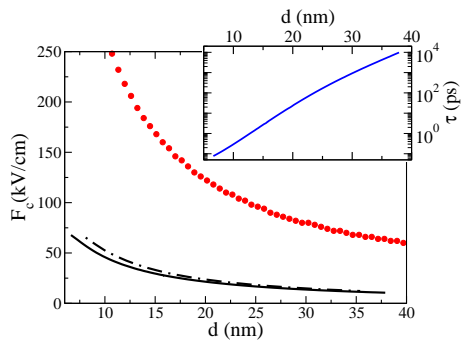}}
\caption{(Color online) (a)  Critical electric field $F^{Te}_{\rm c}$ (circles) required to take one of the two donor electrons to the interface. $F^{Te}_{\rm c}$ depends strongly on the distance $d$ of the donor from the interface. We compare it to the solution to the single donor problem (P) in Ref.~\onlinecite{QC-PRL06} (solid line) where an anisotropic donor wave-function was used. The result for an isotropic donor wave-function for P (dot-dashed line) with $m^*=0.29819 m$ gives a very similar result to the more realistic anisotropic one.
(b) Tunneling time required to ionize an electron on the first excited state of a double donor. This time is the same as required to ionize the single donor P~\cite{QC-PRL06}.}
\label{fig:P-Te}
\end{center}
\end{figure}

We calculate $F^{Te}_{\rm c}$, the electric field required to take one donor electron to the interface, from the condition $\langle \Psi_A | H| \Psi_A \rangle =\langle \Psi_B | H| \Psi_B \rangle$.
In Fig.~\ref{fig:P-Te}(a) we show $F^{Te}_{\rm c}$ versus $d$. The results are similar to the single electron approach~\cite{Kane00PRB} : when the double donor is located a distance $d=20$ nm from the SiO$_2$ interface, $F^{\rm Te}_{\rm c} = 190$ kV/cm.
These values are very large and would probably cause electrical breakdown in the nanostructures. The lower curves in Fig.~\ref{fig:P-Te}(a) correspond to the single donor P results~\cite{QC-PRL06} and are shown for comparison. The curves correspond to isotropic and anisotropic $\Psi_D$ wave-functions, respectively, and are very similar, validating the isotropic approximation adopted here for the double donor problem. In Fig.~\ref{fig:P-Te}(b) we reproduce the tunneling times $\tau_{\rm P}$ required to ionize P~\cite{QC-PRL06}. $\tau$ is proportional to the overlap between states $A$ and $B$, which is orders of magnitude smaller in the double donor problem than in the single donor problem. Therefore, we expect $\tau_{\rm Te}$ to be orders of magnitude larger than $\tau_{\rm P}$ and hence probably of the same order of the spin decoherence time in Si ($\sim 1$ ms in the bulk). Consequently, double donors in the ground state are of no practical use to single spin measurements.

This is not the case if the double donor is in its first excited [1s(A$_1$),1s(T$_2$)] state, which would be more directly accessible for electric field driven (1,1) ${\leftrightarrow}$ (0,2) manipulations. In contrast to the large binding energy and small Bohr radius of the deep centers ground state, the outer electron in the first excited states of neutral double donors has similar properties (binding energy $\sim 32$ meV and Bohr radius $\sim 2$ nm)~\cite{ning71-II} than an electron on a single donor ground state~\cite{Kohn}. Therefore, $F_{\rm c}$ required to singly ionize [1s(A$_1$),1s(T$_2$)] and the tunneling times involved in this process are similar to the ones calculated for P, shown in Fig.~\ref{fig:P-Te} and, therefore, experimentaly meaningful~\cite{QC-PRL06}. We propose a scheme to measure spin dephasing times $T_2^*$ which involves double donors in their first excited state.  As opposed to previous works which have measured spin dephasing times of electrons at donors in bulk Si~\cite{tyryshkin03} and at donors $50$ nm from the interface~\cite{schenkel06}, our proposed experiment would measure the spin dephasing time {\em at the interface}.

In Fig.~\ref{fig:1s-levels-dd} we show the first ionization energies
for the ground state (singlet) and the first excited state (singlet or
triplet) of the neutral double donors. The first excited states have
been observed by looking at the absorption spectra from the ground
state [1s(A$_1$),1s(A$_1$)]~\cite{peale88,bergman88}. Transitions to
pure spin-triplet states are forbidden by selection rules however
triplets have been observed in Te and Se~\cite{bergman88,peale88}: The
spin is not a good quantum number due to spin-orbit interaction. On
the other hand, only one line, due to the spin-singlet state, has been
observed in S, due to the much smaller spin-orbit interaction for the
lighter donor. The spin-triplet-state lifetimes in Te and Se are of
the order of tens of picoseconds as estimated from the linewidths
(which could be limited by experimental precision)~\cite{peale88}, not
much longer than the spin-singlet-state lifetimes. The spin-singlet
lifetime in S is expected to be of the same order as for Te and Se,
while the spin-triplet lifetime must be much longer due to the
relatively small spin-orbit interaction~\cite{atomic-s-o}. Moreover,
it has been recently reported~\cite{karaiskaj03} that the linewidths
of absorption transitions in P-doped Si can be decreased (and,
therefore, the lifetimes increased) by using isotopically purified Si,
while it is expected that the linewidths of deeper impurities (like S)
could be even more dramatically reduced due to the smaller Bohr
radius~\cite{cardona-RMP}. Therefore, in S the first excited triplet
state lifetime is expected to be much longer than the first excited
singlet state lifetime. In what follows, we consider Si doped with S. 

\begin{figure}
\begin{center}
\resizebox{50mm}{!}{\includegraphics{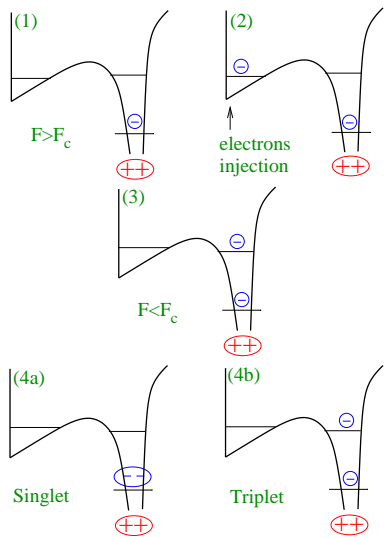}}
\caption{(Color online) Double donors in their first excited state can be used to measure the spin-dephasing times in Si close to an interface. See text for details.
}
\label{fig:dd-proposal}
\end{center}
\end{figure}

For S-doped Si, it
has been observed experimentally that about 50$\%$ of the donors remain
singly ionized~\cite{grimmeis92} (their only electron is in the inner 1s(A$_1$) state) while the others remain neutral and inert for
the proposal presented here. We could use such singly ionized donors to measure the spin dephasing time by manipulating its outer electron between the interface and the sulphur 1s(T$_2$) orbital (under a magnetic field $<1$ T to define a quantization axis) in the way illustrated in Fig.~\ref{fig:dd-proposal}: (1) an electric field slightly larger than the $F_{\rm c}$ required to ionize the donor in its first excited state is applied; (2) electrons are injected at the interface. The ionized donors will bind the new electrons at the interface~\cite{QC-PRL06} leading to a  (1,1) configuration;
(3) the electric field is decreased so that, at $F_{\rm c}$, the electron at the interface goes to the outer 1s(T$_2$) state at the donor [configuration (0,2)] with typical tunneling times as shown in Fig.~\ref{fig:P-Te}(b). (4a) If the two-electron spin state is a singlet, or contains a significant singlet component, the orbital state will rapidly decay to the ground state [1s(A$_1$),1s(A$_1$)], (4b) while if they form a triplet, selection rules imply a much longer lived state. The electric field is increased again so that the outer electron at the donor can go back to the interface. Only those electrons whose spins form a triplet with the inner electron at the donor will remain in the excited state long enough to be able to shuttle [(1,1) $\leftrightarrow$ (0,2)] when the electric field dithers around $F_{\rm c}$. The inner electron remains bound to the donor because the field required to doubly ionize the donor is much larger ($>> F^{Te}_{\rm c}$). Detection of charge at the interface will give a decreasing population of electrons from which we can extract the spin dephasing time of electrons at the Si/SiO$_2$ interface. 

There are various timescales involved in this problem: the unknown spin-triplet lifetime of the first excited state in S (which we have argued should be orders of magnitude longer than the spin-singlet $\sim$ ps lifetime), the tunneling times for the electron shuttling [as given in Fig.~\ref{fig:P-Te}(b)], and the frequency of the electric field dithering (with corresponding times $\gtrsim 1$ ns). 
The proposed experiment would be successful in measuring spin decoherence at the interface if the spin-triplet lifetime is longer than the other two timescales. In this case, such measurements would also give estimates for the donor first excited state spin triplet lifetime. 
It is reasonable to assume that the spin coherence time at the interface can be made much longer than any of the other timescales by improving the interface quality and by isotopically purifying the Si host. Note that the already long bulk value $\sim 1$ ms can be dramatically increased rather easily to $100$ ms or longer by isotopic purification~\cite{witzel05}.

\begin{acknowledgments}
This work is supported by LPS and NSA. B.K. also acknowledges support from CNPq, FUJB, Millenium Institute--MCT, and FAPERJ.
\end{acknowledgments}

\bibliography{double-donor}

\end{document}